# Digital Biosensing By Foundry-Fabricated Graphene Sensors

Brett R Goldsmith*, Lauren Locascio, Yingning Gao, Mitchell Lerner, Amy Walker, Jeremy Lerner, Jayla Kyaw, Angela Shue, Savannah Afsahi, Deng Pan, Jolie Nokes, Francie Barron

## Abstract

Biomedical and environmental testing currently require access to highly specialized facilities or extensive training. Current biochemical tests require complex reagents to achieve a simple human readable result such as a color change. This results in low-information tests, with high reagent costs. To maximize information from testing while minimizing costs, biotechnology should leverage complex analysis enabled by advanced portable computing power and use simplified reagents. Nanotechnologists have created proof-of-concept devices to move diagnosis and monitoring from large core labs to hand-held, easy-to-use personal devices, but none have achieved large-scale manufacturability. We demonstrate the first economical nanoelectronics produced in a commercial foundry. The mass-manufacture of these graphene-based digital biosensors is achieved by successfully integrating delicate graphene material into standard electronic high-volume production processes. We describe critical manufacturing parameters and the results of biosensing measurements that establish graphene biosensors as a portable, low-cost method to gain precise data without labeled, complex reagents. The low power and resource usage of these biosensors enable individuals to gain personal, immediate control over precise biological and environmental data.

## Introduction

The world is entering an inflection point in medical and biological testing with the simultaneous emergence of improved testing technology, advanced software tools, and increased expectations for quality healthcare worldwide. Organizations like the Qualcomm Tricorder XPRIZE and Gates Foundation have pushed for integrations of varied technologies in clinical tests to demonstrate potential. Traditional healthcare companies such as Abbott market point-of-care tools that perform a limited number of assays based on mature assays already cleared by the FDA. In each case, complex, analyte-specific reagents and intricate protocols create a need for multiple platforms and deep biochemical expertise to replicate the capability of a central lab. This only presents an illusion of portability and democratization of biochemical testing. We believe that new label-free measurement tools incorporating graphene sensors will remove the need for liquid reagents, decrease power requirements, and shrink the size of handheld testing devices. These tools will be capable of performing a wide variety of chemical and biochemical assays with a single reader built on a single sensor manufacturing chain, leading to lower cost and standardized protocols.

To demonstrate and validate this approach, we have commercially produced and sold a digital biosensor based on graphene-enabled Field Effect Biosensing (FEB). These sensors are a type of Field Effect Transistor (FET). They can be described as a type of biologically specialized Ion Sensitive FET (ISFET). We describe here the sensing mechanism, demonstrate a label-free capture assay, and summarize the critical manufacturing and quality control milestones met during recent sensor production.

## Background theory:

The unique attributes that are required to build effective field effect biosensors are a combination of semiconductor behavior with chemical stability in air and salt water. Only a few materials, such as graphene, carbon nanotubes, and molybdenum disulfide have this unique combination of chemical stability and electric field sensitivity. This has led to a dense literature covering chemical and biological sensors made using these materials[1–13]. Several attempts were made to produce carbon nanotube biosensors for biomedical use in the early part of the 21$^{st}$ century with frustrating results due to manufacturing difficulties[8]. Fabrication techniques using molybdenum disulfide have not matured sufficiently for devices to move beyond the proof of concept stage[9].

Previously, we have shown that graphene biosensors can be reliably manufactured in research-oriented clean rooms, that the sensors constructed are functional in the hands of independent biologists working in both buffer and serum, and that the performance of the sensors is appropriate for diagnostic development [4,14,15]. In this study, the devices and graphene are manufactured in commercial fabrication facilities under industrial quality control processes. This demonstrates, for the first time, movement away from theoretically "scalable" nanoelectronics techniques to "scaled" manufacturing.

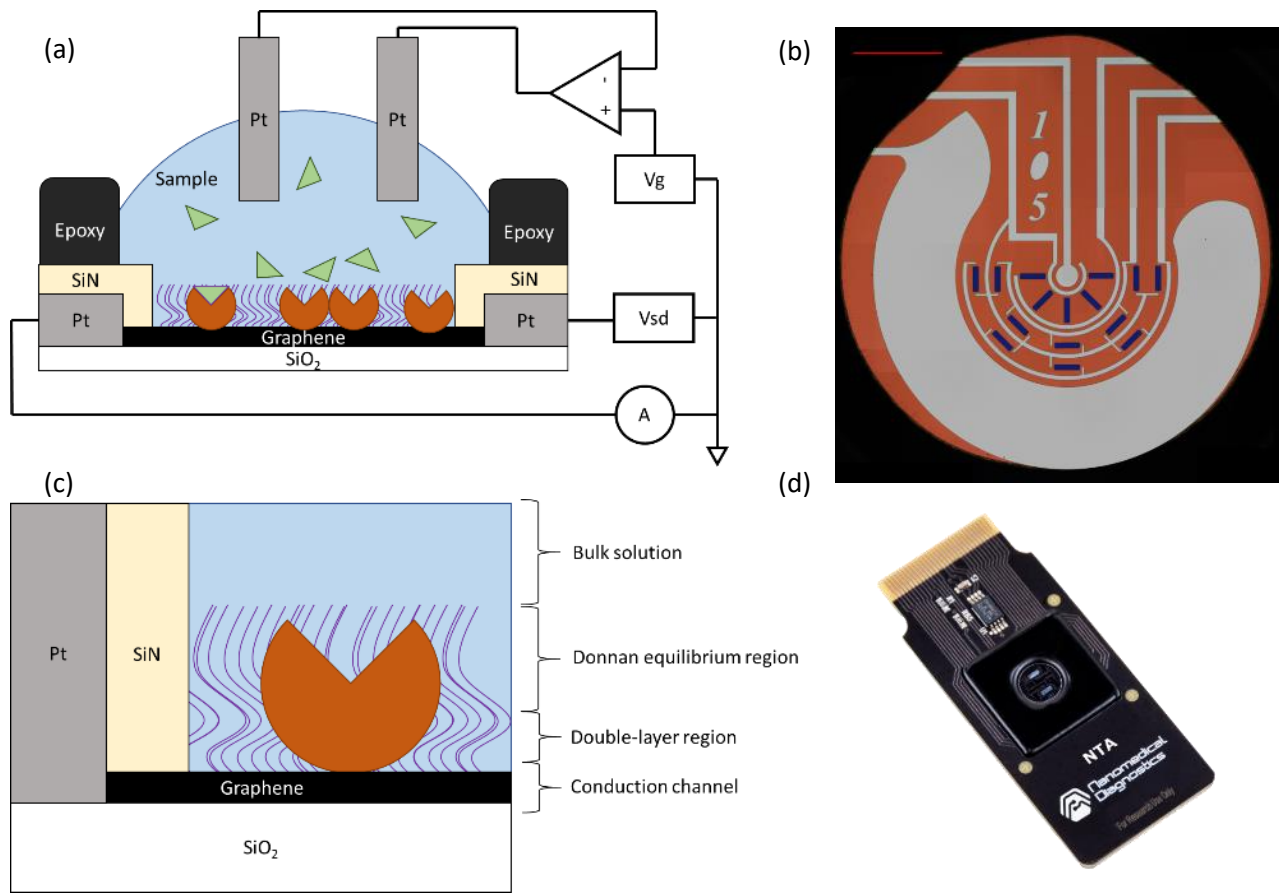

**Figure 1:** (a) Diagram of the sensor architecture. Circular sections on top of the graphene represent proteins embedded in a blocking layer, represented by curved lines. (b) A microscopy image showing an entire sensor surface. Red scalebar is 1 mm. There are fifteen graphene strips grouped into three groups of five, exposed through the silicon nitride encapsulation layer. The center of the circuit is the gate measurement pad (pseudo-reference) and the large pad surrounding the graphene strips is the liquid gate (counter electrode). (c) Diagram of the sensor regions near the graphene. The double layer region is $0.3/\sqrt{c_s}$ nm tall, where $c_s$ is the ionic strength of the bulk solution. The Donnan equilibrium region is the thickness of the combined protein and blocking layer on the surface. (d) Picture of the complete biosensor.

A diagram of our device architecture is shown in Figure 1a, and a top-down microscopy image of the active region of the biosensor is shown in Figure 1b. During measurement, a liquid drop is placed onto the circular region defined by the black epoxy shown here. The platinum counter and reference electrodes built into the sensor surface control and monitor a voltage in the bulk liquid. A blocking layer and embedded biomolecules such as proteins are immobilized onto 15 graphene sensors on the surface. The graphene sensor function relies on several overlapping effects, with a basic cartoon of the relevant layers shown in Figure 1c. A picture of the complete biosensor is shown in Figure 1d.

Closest to the graphene channel, the salts in solution will form a well-organized and well-understood layer on top of the surface countering any difference in charge between the surface and the liquid. The thickness of this layer is the Debye length and is simplified in aqueous solutions to $0.3/\sqrt{c_s}$ in nanometers, with $c_s$ the ionic strength of the solution. The Donnan effect region extends beyond the Debye length, extending through the thickness of an ion-permeable membrane immobilized to the surface. The voltage in the bulk liquid is controlled by conventional electrochemical means. From an electrical perspective, the system can be understood with the bulk liquid as the gate of a transistor, and the combined Donnan region and Debye length as the dielectric between the graphene channel and the gate. From a biological perspective, the system can be understood as a voltage sensitive membrane incorporating proteins with driven voltages, like action potentials, in the bulk liquid. The model below explains how sensing is accomplished for binding interactions, even when charge transfer is not involved.

$$I \approx \frac{W}{L} \mu\, C_g\, V_{sd}\bigl(V_0 - V_g + 2.3\, \phi_{th} \alpha\, \Delta pH + (1-\alpha)\Delta\varphi_D\bigr) \tag{1}$$

Equation 1 shows a modification of previously developed compact models for graphene FETs when combined with ISFET models[16–21]. This model generally relates the current ($I$) to typical electrical properties such as the charge carrier mobility ($\mu$), capacitance per unit area to the gate ($C_g$), width ($W$) and length ($L$) of the graphene channel, source-drain voltage applied directly to the graphene ($V_{sd}$) and the gate voltage ($V_g$) relative to the Dirac voltage ($V_0$). The Dirac voltage here is the gate voltage at which there is a minimum in conduction at neutral pH. The capacitance between the graphene and the liquid, $C_g$, is a series combination of the graphene quantum capacitance, the double layer capacitance, and a capacitance across the immobilized layer due to the Donnan effect[22,23].

The remaining terms are corrections to the gate voltage due to the influence of pH changes and the Donnan potential. This model assumes operation of the sensor near room temperature, for an equivalent gate voltage less than the Dirac voltage, for source-drain voltages below 1 V, and for a channel length greater than 10 μm.

Electronic sensitivity to pH is typically attributed to hydrogen binding to a gate dielectric. For graphene transistors, there is no gate dielectric, but this form of pH sensitivity has been shown to apply through direct shifts in the apparent Dirac voltage[17,24]. The surface pH sensitivity factor ($\alpha$), is a material dependent value. For clean isolated graphene, $\alpha$ is a very low 0.02, but in practice this value is increased by the presence of oxides and nitrides used in device fabrication; $\alpha$ values for practical graphene transistors are around 0.37[25–27]. This term is combined with the thermal voltage ($\phi_{th}$), about 26 mV, and pH shift from a neutral surface ($\Delta pH$) to produce the equivalent gate voltage due to pH.

A Donnan potential ($\Delta\varphi_D$) is created when an ion-permeable layer separates 2 collections of ions, as shown in equation 2.

$$\Delta\varphi_D = \phi_{th}\, ln\frac{\left(\sqrt{4c_s^2 + c_x^2} + c_x\right)}{2\, c_s} \tag{2}$$

In this case, the immobilized layer of proteins, peptides, surfactants, PEG, or other soft molecules separate the graphene channel and double-layer from the bulk solution. Any charges or dipoles leading to a net charge within that immobilized ion permeable layer ($c_x$) will require an extra accumulation of a counter-ion within the layer to maintain charge neutrality. This difference in the concentration of ions between the bulk solution ($c_s$) and that in the immobilized protein layer creates a Donnan potential[20,28]. This additional potential enables sensing beyond the Debye screening length,[19,22,29] and has been demonstrated repeatedly with graphene biosensors[4,7,12,14,30].

## Results

### Chemical Measurements

This model provides a relatively simple framework for thinking about how a graphene-based digital biosensor works. Practical function of the biosensor is demonstrated by performing sensing measurements. We start by evaluating the basic response of the biosensors to common background effects such as shifts in pH and ionic strength.

Our sensor data is analyzed by calculating the percent change in current from a baseline taken in assay buffer. This removes the effect of variations in resistance sensor-to-sensor. Additionally, the change in current relative to a controlled change in gate voltage $\Delta I/\Delta V_g$ is used to isolate responses due to change in gate capacitance $C_g$, independent of direct potential shifts from pH and Donnan potential. To generate the data in the Figure 2, the standard Agile R100 measurement settings were used, which limit applied voltages to between +/- 100 mV[4,14]. The chip fabrication process purposefully creates a shift of the Dirac voltage to greater than 100 mV, so that $\Delta I/\Delta V_g$ is approximately linear over all applied $V_g$. This simplifies and speeds up practical biosensor measurement and analysis, while unfortunately preventing direct measurements of the Dirac voltage.

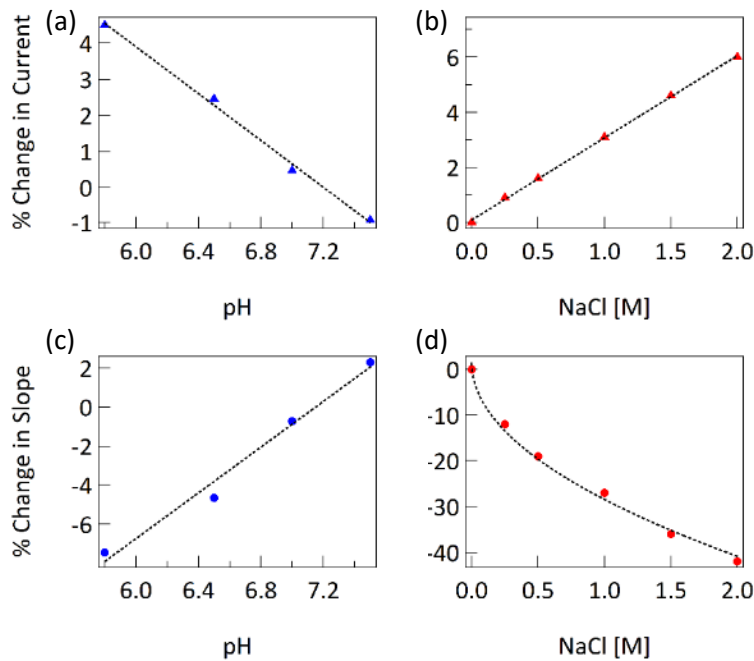

**Figure 2:** (a) Change in current due to change in pH. Fit is linear with a slope of -3.2% per pH unit. (b) Change in current due to change in ionic strength, when pH is held constant. Fit is linear with a slope of 3.0% per molar unit of NaCl. (c) Change in slope (dI/dVg) due to change in pH. Fit is linear with a slope of 5.9% per pH unit. (d) Change in slope due to change in ionic strength. Response is fit to $-0.3\sqrt{[NaCl]}$

The solutions used in Figure 2a and 2c are standard 1X phosphate buffered saline (PBS) pH 7.4 with varying amounts of 100 mM HCl added to adjust the pH. The pH value was set first at pH 7, and then measurements were performed at increasing pH values. The solutions used in Figure 2b and 2d are NaCl in deionized water.

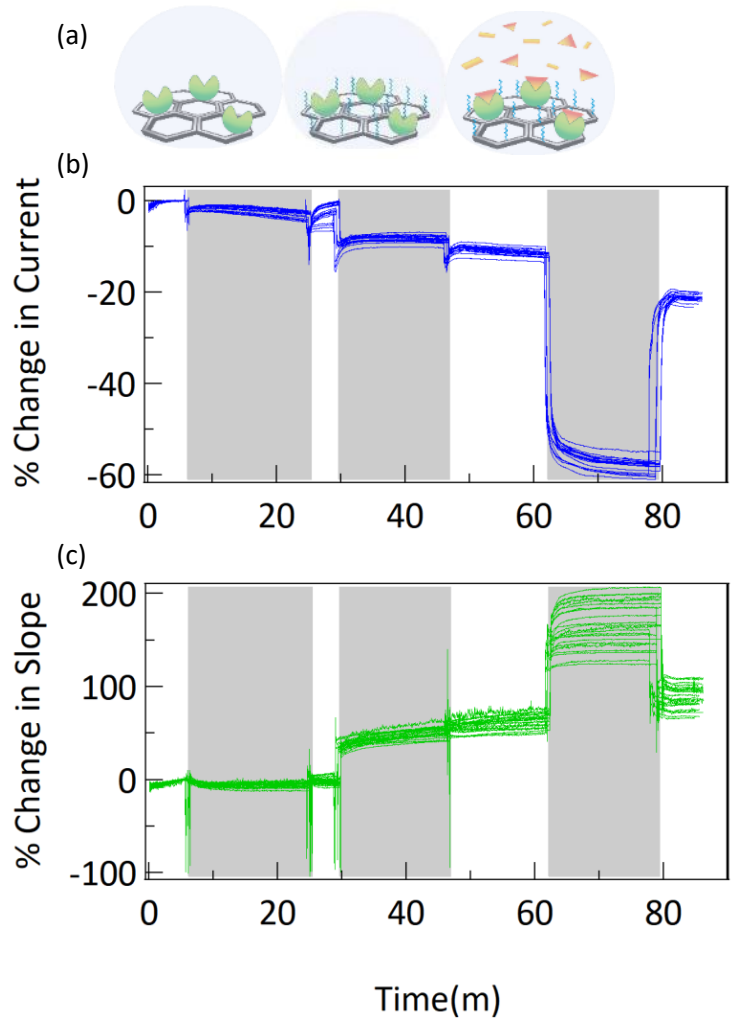

**Figure 3:** (a) Diagram of the steps of protein immobilization and measurement used here. First, antibodies against IL-6 are immobilized, then PEG is added as a block, then measurements are performed with IL-6. (b) Change in current and (c) change in slope for 23 different sensors during immobilization process. Shading is used to delineate steps: calibration in MES buffer, COOH activation by EDC/NHS incubation in MES buffer, wash in MES, antibody incubation in PBS, PEG incubation in PBS, quench in ethanolamine, wash in PBS.

The most notable feature of the data in Figure 2 is the dependence of the gate response slope on the square root of the ion concentration. In Equation 1, both $\alpha$ and $C_g$ are dependent on the ion concentration, but the double layer capacitance component of $C_g$ is simplified to $\varepsilon \sqrt{c_s}/0.3$. The observed and modeled dependency of these sensors on pH and ionic strength indicates an ability to generate broad and complex responses, capable of creating signatures for diverse biological macromolecules and their complexes.

## Protein Measurements

To demonstrate measurement of a biological interaction, a monoclonal antibody against human interleukin-6 (anti-IL6) and recombinant human IL-6 (IL6) were used from a commercial ELISA kit. Anti-IL6 was immobilized on graphene chips with a prepared carboxyl surface and activated via EDC/sNHS (Figure 3), as described previously[4]. Timing for each of the liquid exchange steps is described in the methods section. All liquid exchanges are done via an aspirator and manual pipette. Aspiration of liquid from the sensor surface briefly removes the liquid gate, prior to re-establishing the gate with pipetting of new liquid to the surface. This leads to the spikes in the data at the white/grey junctions.

Figure 3b and 3c show the percent change in current and slope responses for 23 different sensors all following the same immobilization protocol with the same anti-IL6 reagent. The immobilization data shows consistent trends. Addition of EDC and sNHS after calibration always leads to a decrease in current, without a change in slope. Addition of the antibody at 30 minutes leads to a further decrease in current and a large increase in slope. This change is likely due to switching from the pH 6.0 MES buffer used during carboxyl activation to the pH 7.4 PBS pH buffer used during antibody incubation. Addition of PEG-amine after anti-IL6 immobilization always leads to a small decrease in current and a small increase in slope. This is consistent with association of PEG to the surface to serve as a blocking layer around immobilized protein. Addition of ethanolamine (pH 8.0) quenches any remaining activated carboxyl groups, and always causes a large decrease in current and a large increase in slope. Rinsing with PBS (pH 7.4) then raises the current and decreases the slope, although never back to the initial starting position. This repeatable set of chemistry and sensor responses indicates both reproducibility of response sensor-to-sensor as well as demonstrable, permanent change to the surface chemistry of the chips due to the immobilization process. The sign of the responses we see during immobilization are consistent with the expected responses from the pH changes measured on bare graphene chips shown in Figure 2. However, the magnitude of responses is significantly higher. As pyrene, antibody, and PEG are added to the chip, a layer supporting a Donnan effect voltage is created.

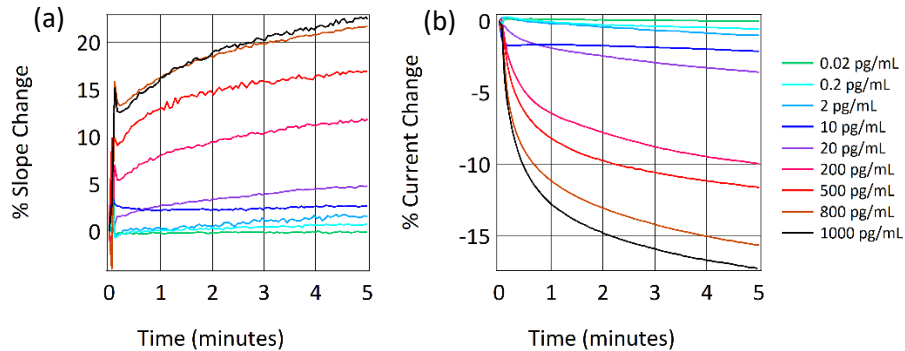

**Figure 4:** (a) Change in gate slope of sensors functionalized with antibodies against IL-6 to different concentrations of IL6. (b) Change in current for the same sensors.

The effectiveness of this immobilization process is demonstrated in the sensing measurements shown in Figure 4. Different concentrations of IL-6 in PBS were applied to the anti-IL6 immobilized chips. The different concentrations of IL6 lead to different response magnitudes and different speeds of interaction, as expected by any kinetic binding measurement. This data compares well with the metrics provided by the ELISA kit from which the reagents were taken. When running an ELISA with this kit, the expected sensitivity is <2 pg/mL (FEB sensitivity is <2 pg/mL), and the standard range of response is between 7.8 pg/mL and 500 pg/mL (FEB range of response is 2 pg/mL to 1000 pg/mL). Here, we have reproduced the ELISA quality while adding kinetics. We have removed the need for labels while providing more information-dense real-time data than an end-point assay technique like ELISA.

## Scaled Manufacturing

The data shown here establishes that the devices we have manufactured are similar in function to other biosensors in the graphene literature. A major hurdle to providing graphene biosensors to biological researchers is the typical manufacture of nanoelectronics in research-oriented clean rooms by teams of research scientists. Previously, there was no nanoelectronic manufacturing approaching ISO9001 standards or Good Manufacturing Practices (GMP), despite those requirements for commercial and clinical applications.

The facilities used to manufacture chips are located at Rogue Valley Microelectronics, Nanomedical Diagnostics ("Nanomed"), and Samtec. Rogue Valley used a 150 mm MEMS processing line, without modification to staff or equipment to process graphene wafers. The Samtec facility similarly used established silicon chip packaging equipment, staff, and traceable processes to package these graphene chips.

Graphene is grown and transferred at Nanomed using automation of a previously described process[4]. The marginal cost for graphene, when using an automated system such as ours, is $0.019 per cm$^2$. This comes from our average cost of copper foil of $0.012 per cm$^2$ and the cost of power to run the furnaces, which in California is currently $0.007 per cm$^2$ of grown graphene. This is less than the price of silicon wafer substrates for this work, $0.40 per cm$^2$ for 150 mm wafers. The cost of processing wafers is 20x greater than the cost of the graphene raw material, making the processing the dominant cost factor in producing a commercial graphene device and not the graphene raw material cost.

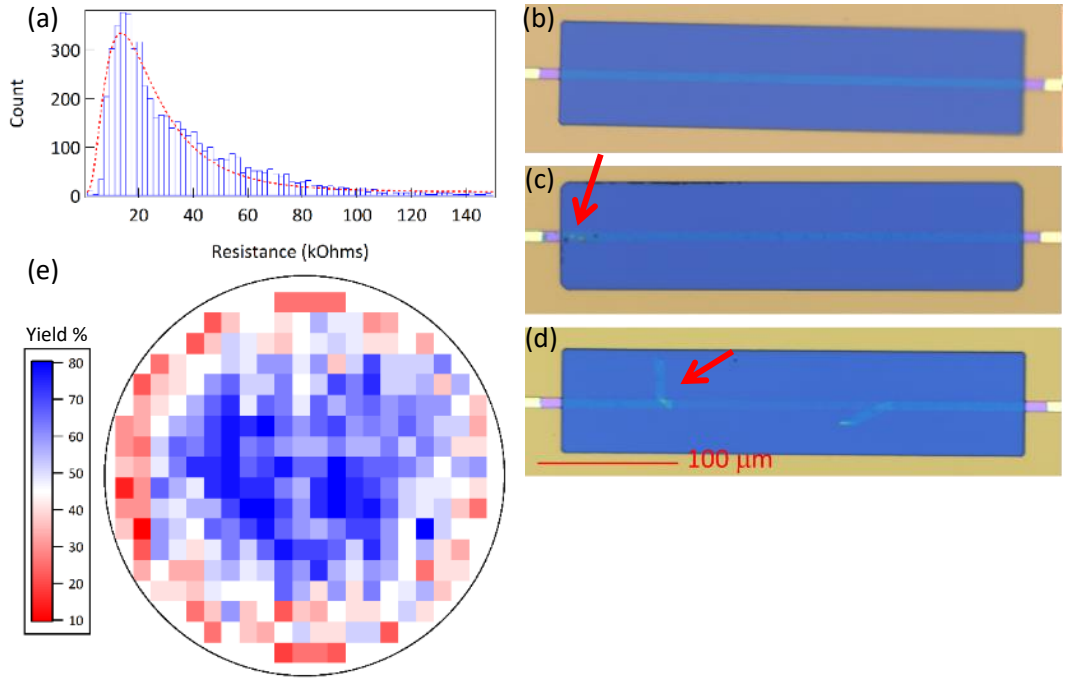

**Figure 5:** (a) Histogram of test resistances from 5543 chips. Fit is a log-normal distribution with a peak at 13.6 kOhm and a standard deviation of 9.2 kOhm. (b) Microscopy image of defect-free graphene sensor (c) Microscopy image of a graphene sensor with minor polymer contamination highlighted by the red arrow. (d) Microscopy image of a graphene sensor with major graphene tearing highlighted by the red arrow. (e) Wafer yield map combining data from 27 wafers, showing cumulative % yield for each die.

Our process for manufacturing digital biosensors starts with deposition and patterning of a metallization layer that creates the routing for the source-drain voltage on 15 graphene transistors per die, as well as the platinum reference and counter electrodes. Graphene is grown and transferred onto the wafer. After deposition, graphene is etched using a hard mask and an oxygen plasma. A silicon nitride encapsulation layer is deposited via PECVD. Windows are etched into the silicon nitride down to the graphene hard mask. Wafers are diced and packaged via a chip-on-board process, where an epoxy is used to protect wirebonds and define the liquid wetted area. After packaging, the chips are annealed and undergo strict QA processes to ensure reproducible quality.

Figure 5 presents quality assurance data from three production lots comprising twenty seven 150 mm wafers and 7,992 chips produced over the course of one year. Quality control processes focus on reproducibility and cost rather than searching for occasional outstanding performance. Specifically, this means that we use automated optical microscopy to search for contamination and graphene tearing over entire wafers, rather than using nanotechnology specialized hardware such as atomic force microscopy or Raman spectroscopy. Differential interference contrast microscopy is used along with AI-enabled defect identification to perform optical evaluations, such as those shown in Figure 5. Table 1 shows the distribution of defects found via optical microscopy across all 27 wafers. Minor defects generally do not prevent a chip from passing to the next stage of manufacture, while major defects always prevent a chip from continuing.  A chip may have no defects, a single defect, or multiple defects.

*Table 1 Defect frequency.*

| Issue | number | Percent of Errors |
|---|---|---|
| Minor Polymer Contamination | 4,066 | 44.2% |
| Minor Tears | 2,857 | 31.0% |
| Major Tears | 1,742 | 18.9% |
| Major Polymer Contamination | 295 | 3.2% |
| Major Lithography Defect | 147 | 1.6% |
| Minor Lithography Defect | 77 | 0.8% |
| Other | 22 | 0.2% |

After optical QA, electrical QA is performed. Figure 5a shows a histogram of resistances from 5,543 chips. This data can be fit with a log normal distribution, with a peak of 13.6 kΩ ± 9.2 kΩ. The length of the measured graphene strips is 300 μm, with a width of 10 μm. Five strips are measured in parallel, for a typical resistance/sq of 2.3 kΩ/sq. Transistors that have resistances outside a range from 1 kΩ to 100 kΩ fail electrical QA. This bounds the power draw per sensor channel at between 1 nW to 100 nW when using our standard 10 mV $V_{sd}$. This low power draw anticipates the need for thousands of chemically differentiated sensor channels or other graphene transistors in a portable form factor.

At the end of the QA process, a map of yielded dies on the wafer is produced. Combining maps from 27 wafers shows a clear bias for failure at the edge of the wafers, probably indicating damage to the graphene transistors or lithography due to handling or normal edge effects. For example, major lithography defects occur predominantly at the wafer edge. Together the data in Figure 5e shows a percent yield by die position, with an overall average of 52%, a value that compares well with silicon device yields in the early 1980s[31].

## Discussion

For the past year, we have produced and sold graphene sensors for biological researchers following these described processes. We present the first complete characterization of this fusion of electronics and biochemistry. This marks a break from the past few decades of great feats of artisanal nanoelectronics and demonstrates the potential for graphene biosensors to serve as the interface between biochemistry and digital analysis.

A core challenge has been making the transition from an expensive, variable sensor to a cost-efficient, consistent batch-to-batch manufactured sensor. A key to that transition is following a rigorous specification and quality control process appropriate for a biologist end user rather than a nanotechnologist. While the manufacturing and QA processes described here were used to create graphene-based biosensors, these processes are broadly enabling for many electronics applications using graphene and other nanomaterials.

The demonstrated biological sensing capability, low power requirements, and compact size of graphene-based biosensors will enable development of the next generation of biochemical applications. With the most difficult piece of the puzzle – cost-effective large-scale manufacturing – solved, low-power, portable digital biosensors can significantly impact healthcare industries with innovative new products that enable cutting-edge life science research, drug discovery applications, and diagnostic and health monitoring platforms.

## Methods

The Agile R100 system was used for all measurements. The standard electrical settings were used. The gate voltage was swept between ± 100 mV in a triangle wave at a slow speed of 0.3 Hz, while $V_{sd}$ was held at 10 mV. These voltage ranges were selected to minimize the electric fields on the proteins. Agile Plus software was used to run the hardware.

Except where noted, 1X phosphate buffered saline solution pH 7.4 (PBS) (ThermoFisher # 10010031) was used for calibration and measurement. HCl was used to create PBS solutions of varying pH from 5.8 to 7.5, measured with a calibrated glass electrode pH meter. Fifty millimolar 2-(N-morpholino)ethanesulfonic acid buffer pH 6.0 (MES) (Alfa Aesar # J62574-AK, diluted in DI water) was used for crosslinking chemistry. EDC (1-ethyl-3-(3-dimethylaminopropyl)carbodiimide hydrochloride) (Amresco # N195) was used with sNHS (N-hydroxysulfosuccinimide) (G Biosciences # BC97) to activate carboxyl groups for amine attachment. For ionic strength measurements, deionized water was used for calibration, and concentrations of NaCl from 0.25 M to 2 M in deionized water were used for the measurement. Agile FLEX biosensor chips (Nanomed) were used for pH and ionic strength measurements.

Anti-IL6 and IL6 were purchased as part of an ELISA kit (ThermoFisher # KHC0062).

Agile COOH biosensor chips (Nanomed) were used for IL6 measurements. COOH chips are prepared via incubation of clean graphene chips with 3mM pyrene-carboxylic acid (TCI # P1687) in methanol for two hours.

The detailed process for anti-IL6 immobilization is: After calibration in 50 mM MES pH 6.0 for 5 minutes, 2.08 mM EDC and 5.53 mM sNHS in 50 mM MES pH 6.0 was incubated for 20 minutes. The chips were rinsed 2 times with 50 mM MES pH 6.0, followed by 14.6 nM anti-IL-6 in 1X PBS pH 7.4 for 15 minutes. Then, 3 mM PEG-amine (Broadpharm Item #BP-22355) in PBS was incubated for 15 minutes to block the surface, followed by 1 M ethanolamine pH 8.0 (Alfa Aesar Cat# L14322) for 15 minutes to deactivate any remaining carboxyl groups. Finally, the biosensor chips were rinsed 5 times in PBS, and the last rinse was incubated for 5 minutes to stabilize the signal prior to measurement. This process was performed with a hand pipette following software prompts.

Prime Si, 150mm, P-type wafers were purchased from Rogue Valley Microdevices, as was all mask fabrication and lithography. A wet thermal oxide of 3,000 Å was first grown. A metallization layer of 100 Å Cr and 500 Å Pt was patterned via liftoff.

Graphene was grown via chemical vapor deposition at Nanomed on copper and transferred to the wafer via bubble transfer. Microscopy was used to evaluate graphene quality at every transistor location prior to further processing.

At Rogue Valley Microdevices, a 1,000 Å Au layer was deposited on the graphene and etched to form a hard mask. An oxygen plasma etch was used to remove excess graphene. A 5,000 Å SiN layer was deposited via PECVD, patterned and etched via RIE.

Custom PCBs for packaging were designed by Varasco Engineering, laid out by Pacific Design, and purchased from Consisys. Samtec diced wafers after microfabrication and performed a standard chip-on-board packaging process using gold wirebonds and a custom dam-and-fill encapsulation pattern to create the liquid well on the sensor.

The chips are cleaned, vacuum annealed at 200 C, and inspected at Nanomed.

Optical microscopy is performed using an nSPEC microscope (Nanotronics). AI driven automated defect identification software is also provided by Nanotronics.

Electrical QA is performed using an Agile R100 (Nanomed) in QA mode.

The datasets generated during and/or analysed during the current study are available from the corresponding author on reasonable request.

## Acknowledgments
We would like to thank Hector Aldaz for helpful conversations and manuscript editing. We would like to thank Amanda Zimmer for graphics assistance in Figure 3a. We would like to thank David Giegel for much helpful advice and chairing our Scientific Advisory board.

## Author contributions statement
BG wrote the main text. ML and BG prepared Figure 1. BG and LL prepared Figure 2. LL, SA, JN, AW, and BG prepared Figures 3 and 4. ML, JL, YG, JK, and BG prepared Figure 5 and Table 1. DP wrote software and designed the chips. FB oversaw work by AW, LL, and SA. ML oversaw work by YG, JL, and JK. AS oversaw work by JN.

## Competing Interest statement
All authors are recently or presently employed by Nanomed, with some stock or stock options in the company. All work here was funded by Nanomed.

## References
1. Ohno, Y., Maehashi, K., Yamashiro, Y. & Matsumoto, K. Electrolyte-Gated Graphene Field-Effect Transistors for Detecting pH and Protein Adsorption. *Nano Lett.* **9,** 3318–3322 (2009).

2. Lu, Y. *et al.* Graphene-protein bioelectronic devices with wavelength-dependent photoresponse. *Appl. Phys. Lett.* **100,** (2012).

3. Lerner, M. B., Dailey, J., Goldsmith, B. R., Brisson, D. & Charlie Johnson, A. T. Detecting Lyme disease using antibody-functionalized single-walled carbon nanotube transistors. *Biosens. Bioelectron.* **45,** 163–167 (2013).

4.  Lerner, M. B. *et al.* Large scale commercial fabrication of high quality graphene-based assays for biomolecule detection. *Sensors Actuators, B Chem.* (2016). doi:10.1016/j.snb.2016.09.137

5.  Choi, Y. *et al.* Single-Molecule Lysozyme Dynamics Monitored by an Electronic Circuit. *Science (80-. ).* **335,** 319–324 (2012).

6.  Cohen-Karni, T., Qing, Q., Li, Q., Fang, Y. & Lieber, C. M. Graphene and nanowire transistors for cellular interfaces and electrical recording. *Nano Lett.* **10,** 1098–1102 (2010).

7.  Gao, N. *et al.* Specific detection of biomolecules in physiological solutions using graphene transistor biosensors. *Proc. Natl. Acad. Sci. U. S. A.* **113,** 14633–14638 (2016).

8.  Allen, B. L., Kichambare, P. D. & Star, A. Carbon nanotube field-effect-transistor-based biosensors. *Adv. Mater.* **19,** 1439–1451 (2007).

9.  Sarkar, D. *et al.* MoS2 field-effect transistor for next-generation label-free biosensors. *ACS Nano* **8,** 3992–4003 (2014).

10. Zuccaro, L. *et al.* Real-Time Label-Free Direct Electronic Monitoring of Topoisomerase Enzyme Binding Kinetics on Graphene. *ACS Nano* **9,** 11166–11176 (2015).

11. Xu, S. *et al.* Real-time reliable determination of binding kinetics of DNA hybridization using a multi-channel graphene biosensor. *Nat. Commun.* **8,** 1–10 (2017).

12. Ang, P. K. *et al.* A Bioelectronic Platform Using a Graphene− Lipid Bilayer Interface. *ACS Nano* **4,** 7387–7394 (2010).

13. Kempaiah, R., Chung, A. & Maheshwari, V. Graphene as cellular interface: Electromechanical coupling with cells. *ACS Nano* **5,** 6025–6031 (2011).

14. Afsahi, S. *et al.* Novel graphene-based biosensor for early detection of Zika virus infection. *Biosens. Bioelectron.* **100,** (2018).

15. Qvit, N., Disatnik, M.-H., Sho, J. & Mochly-Rosen, D. Selective phosphorylation inhibitor of δPKC-PDK protein-protein inter-actions; application for myocardial injury in vivo. *J. Am. Chem. Soc.* jacs.6b02724 (2016). doi:10.1021/jacs.6b02724

16. Wang, H., Hsu, A., Kong, J., Antoniadis, D. A. & Palacios, T. Compact virtual-source currentvoltage model for top-and back-gated graphene field-effect transistors. *IEEE Trans. Electron Devices* **58,** 1523–1533 (2011).

17. Mackin, C. & Palacios, T. Large-scale sensor systems based on graphene electrolyte-gated field-effect transistors. *Analyst* (2016). doi:10.1039/c5an02328a

18. Bergveld, P. Thirty years of ISFETOLOGY. *Sensors Actuators B Chem.* **88,** 1–20 (2003).

19. Kaisti, M. Detection principles of biological and chemical FET sensors. *Biosens. Bioelectron.* **98,** 437–448 (2017).

20. Bergveld, P. A critical evaluation of direct electrical protein detection methods. *Biosens. Bioelectron.* **6,** 55–72 (1991).

21. Zuccaro, L., Krieg, J., Desideri, A., Kern, K. & Balasubramanian, K. Tuning the isoelectric point of graphene by electrochemical functionalization. *Sci. Rep.* **5,** 1–13 (2015).

22. Palazzo, G. *et al.* Detection Beyond Debye's Length with an Electrolyte-Gated Organic Field-Effect Transistor. *Adv. Mater.* **27,** 911–916 (2014).

23. Das, S. Explicit interrelationship between Donnan and surface potentials and explicit quantification of capacitance of charged soft interfaces with pH-dependent charge density. *Colloids Surfaces A Physicochem. Eng. Asp.* **462,** 69–74 (2014).

24. MacKin, C. *et al.* A current-voltage model for graphene electrolyte-gated field-effect transistors. *IEEE Trans. Electron Devices* **61,** 3971–3977 (2014).

25. Fu, W. *et al.* Graphene transistors are insensitive to pH changes in solution. *Nano Lett.* **11,** 3597–3600 (2011).

26. Wang, Y. Y. & Burke, P. J. A large-area and contamination-free graphene transistor for liquid-gated sensing applications. *Appl. Phys. Lett.* **103,** (2013).

27. Mailly-Giacchetti, B. *et al.* PH sensing properties of graphene solution-gated field-effect transistors. *J. Appl. Phys.* **114,** (2013).

28. Ohshima, H. & Ohki, S. Donnan potential and surface potential of a charged membrane. *Biophys. J.* **47,** 673–678 (1985).

29. Schasfoort, R. B. M., Bergveld, P., Kooyman, R. P. H. & Greve, J. Possibilities and limitations of direct detection of protein charges by means of an immunological field-effect transistor. *Anal. Chim. Acta* **238,** 323–329 (1990).

30. Gao, N. *et al.* General strategy for biodetection in high ionic strength solutions using transistor-based nanoelectronic sensors. *Nano Lett.* **15,** 2143–2148 (2015).

31. MacK, C. A. Fifty years of Moore's law. *IEEE Trans. Semicond. Manuf.* **24,** 202–207 (2011).